\documentclass{aa}
\usepackage{txfonts}
\usepackage{graphicx}
\usepackage{natbib}
\bibpunct{(}{)}{;}{a}{}{,}
\begin{document}

\newcommand{\iau}{Int.Astron.U.}
\newcommand{\asj}{Astron.Soc.Jap.}

\title{Explosions of O-Ne-Mg cores, the Crab supernova, and subluminous Type II-P supernovae}
\titlerunning{Explosions of O-Ne-Mg cores and the Crab supernova}

\author{F.S.~Kitaura, H.-Th.~Janka, W.~Hillebrandt}
\authorrunning{F.S.~Kitaura et.~al.}
\institute{Max-Planck-Institut f\"{u}r Astrophysik,
Karl-Schwarzschild-Str.1, Postfach 1317, 85741 Garching, 
Germany} 
\offprints{H.-Th.~Janka, \email{thj@mpa-garching.mpg.de}}
 \institute{Max-Planck-Institut f\"ur Astrophysik, 
              Karl-Schwarzschild-Str.\ 1, D-85741 Garching, Germany
              }
\date{\today}

\abstract{We present results of simulations of stellar collapse and
explosions in spherical symmetry for progenitor stars
in the 8--$10\,M_\odot$ range with an O-Ne-Mg core. The simulations were
continued until nearly one second after core bounce and were performed
with the \textsc{Prometheus/Vertex} code with a variable Eddington factor solver
for the neutrino transport, including a state-of-the-art treatment of
neutrino-matter interactions. Particular effort was made to implement
nuclear burning and electron capture rates with sufficient accuracy to
ensure a smooth continuation, without transients, from the progenitor
evolution to core collapse. Using two different nuclear equations of
state (EoSs), a soft version of the Lattimer \& Swesty EoS and
the significantly stiffer Wolff \& Hillebrandt EoS, we found no
prompt explosions, but instead delayed explosions, powered by neutrino heating
and the
neutrino-driven baryonic wind which sets in about 200$\,$ms
after bounce. The models eject little nickel ($< 0.015M_\odot$),
explode with an energy of $\ga0.1\times 10^{51}\,$erg, and
leave behind neutron stars (NSs) with a baryonic mass near $1.36\,M_\odot$.
Different from previous models of such explosions, the ejecta during
the first second have a proton-to-baryon ratio of $Y_{\rm{e}} \ga 0.46$, which suggests a chemical
composition that is not in conflict with galactic abundances. No
low-entropy matter with $Y_{\rm{e}} \ll 0.5$ is ejected. This excludes such
explosions as sites of a low-entropy r-process. The low explosion
energy and nucleosynthetic implications are compatible with the 
observed properties of the Crab supernova,
and the small nickel mass supports the possibility that our models
explain some subluminous Type II-P supernovae.

}

\maketitle

\section{Introduction}
Recent observations of subluminous Type II-P supernovae (SNe) like 
2005cs, 2003gd, 1999br and
1997D, have renewed attention to stars near the lower end of the mass range of
core-collapse SN progenitors, i.e. to stars with about 8--$10\,M_\odot$, which develop
O-Ne-Mg cores. A possible link between both has been suggested because of the 
low $^{56}$Ni and $^{16}$O ejecta masses and low progenitor
luminosities (e.g., Chugai \& Utrobin 2000, Hendry et al.\ 2005). 
However, due to many uncertainties this connection is far from being clear
(e.g., Pastorello et al.~2004, 2005; Hamuy 2003; Zampieri et al. 2003,
and refs.\ therein). Also the
Crab Nebula's progenitor was proposed to be in this mass window (Gott et
al.~1970, Arnett 1975, Woosley et al.~1980, Hillebrandt 1982). The observed composition of the Crab remnant
(small C and O abundances, He overabundance) was interpreted as a strong
indication that the Crab Nebula comes from a collapsing and exploding
progenitor with an O-Ne-Mg core (Davidson et al.~1982, Nomoto et al.~1982, Nomoto 1983).

Moreover, these stars were considered as possible sites for a low entropy r-process
(for example, Hillebrandt 1978, Wheeler et al.~1997, Sumiyoshi et al.~2001, Wanajo et al.~2003)
based on the assumption that they explode by the prompt bounce-shock mechanism, which Hillebrandt et
al.~(1984) found to work in a numerical simulation, taking Nomoto's O-Ne-Mg core
model (Nomoto 1984, 1987). Such explosions are characterised by the direct propagation of the shock
out of the core, the formation of a mass cut, and the continuous acceleration
of the material outside of the mass cut to high velocities. They are expected to
eject relatively large amounts of neutron-rich matter with low $Y_{\rm{e}}$
($\sim$0.2) and low entropies ($\sim\,$10$k_{\rm{B}}$ per nucleon).
However, several groups could not confirm the viability of the prompt explosion
mechanism (Burrows \&
Lattimer 1985, Baron et al.~1987, Mayle \& Wilson 1988). Mayle \& Wilson (1988) continued their simulations in the
post-bounce phase for a longer time and obtained instead a so-called neutrino-driven, delayed explosion (Bethe \& Wilson 1985) with a low production of $^{56}$Ni
(approximately 0.002 $M_\odot$) in agreement with subluminous Type II-P
SNe as mentioned above,
but with a vast overproduction of neutron-rich material (at least 0.02
$M_\odot$ of ejecta with $Y_{\rm e}\la0.41$). The latter finding is inconsistent
with the chemical composition of our galaxy, which allows for no more than 10$^{-3}$ $M_\odot$ of
material with $Y_{\rm e}<0.42$ being ejected per SN (Hartmann et al.~1985).
Moreover, the explosion energies of both studies, around $2\times10^{51}$erg in
Hillebrandt et al.'s (1984) model and between 0.6 and 1.2 $\times10^{51}$erg in
Mayle \& Wilson's (1988) simulations, would be inconsistent with the long
plateau phase of the above mentioned subluminous SNe, if their H-envelope masses were $\la8M_\odot$.  

It was suggested that the reason for the discrepant results in the
SN simulations of O-Ne-Mg cores (prompt explosions, delayed explosions,
no explosions) could be explained by
the different nuclear EoSs used by the groups (Fryer et al.~1999). Having in mind that the different approximations in the neutrino
transport in previous calculations introduced additional uncertainties,
we revisit this topic with a state-of-the-art neutrino transport
treatment together with a careful description of weak interactions and
including relevant nuclear burning reactions. We additionally make a comparison of collapse and post-bounce calculations with different nuclear EoSs.

\section{Numerical techniques and input physics}
 The transport of neutrinos and antineutrinos of all flavors is done with the
 energy-dependent solver for the coupled set of moments equations and
 Boltzmann equation called \textsc{Vertex}. It is described in detail in Rampp \& Janka
 (2002). The equations of hydrodynamics are integrated with the Newtonian
 finite-volume code \textsc{Prometheus}, which uses a third-order, time-explicit Godunov
 scheme. This code is a direct implementation of the Piecewise Parabolic Method
 (PPM), based on a Riemann solver. General relativistic gravity is taken into account approximately by an ``effective
relativistic potential'' according to Marek et al.~(2005). Gravitational redshift and time
dilation effects are included in the neutrino transport (see Rampp \& Janka 2002). 

% Fig: 
\begin{figure}[!t]{
\resizebox{8.7cm}{!}{
\includegraphics{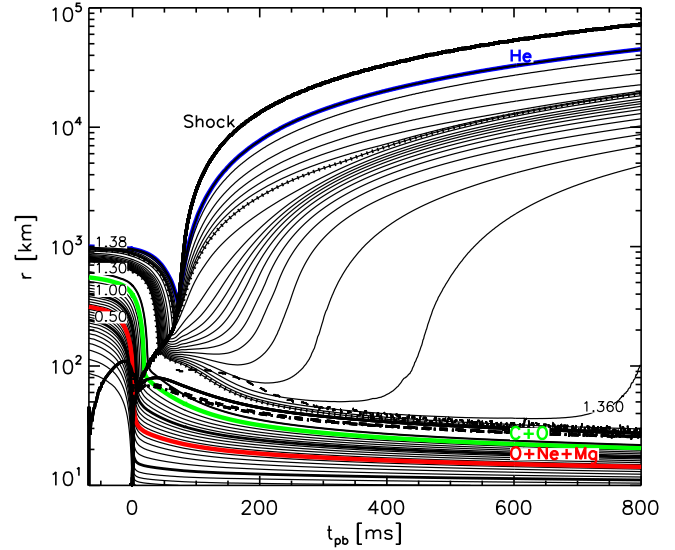}}}
\caption{\footnotesize{Mass trajectories for the simulation with the W\&H EoS as a function of
    post-bounce time (t$_{\rm{pb}}$). Also plotted: shock position (thick solid
  line starting at time zero and rising to the upper right corner), gain radius
  (thin dashed line), and neutrinospheres ($\nu_{\rm{e}}$: thick solid; $\bar{\nu}_{\rm{e}}$: thick
  dashed; $\nu_\mu$, $\bar{\nu}_\mu$, $\nu_\tau$, $\bar{\nu}_\tau$: thick
 dash-dotted). In addition, the composition
 interfaces are plotted with different bold, labelled lines: the inner boundaries of the
  O-Ne-Mg layer at $\sim$0.77 $M_\odot$, of the C-O layer at $\sim$1.26
  $M_\odot$, and of the He layer at 1.3769 $M_\odot$. The two dotted lines represent the
 mass shells where the mass spacing between the plotted trajectories changes. 
An equidistant spacing of $5\times10^{-2} M_\odot$ was chosen up to $1.3579
  M_\odot$, between that value and $1.3765 M_\odot$ it was $1.3\times10^{-3}
  M_\odot$, and $8\times10^{-5} M_\odot$ outside.
}}\label{masW}
\end{figure}

% Fig: 
\begin{figure}[!t]{
\resizebox{8.7cm}{!}{
\includegraphics{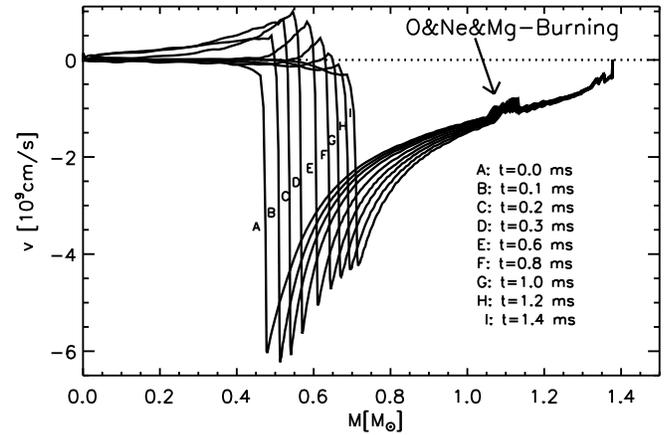}}}
\caption{\footnotesize{Velocity profiles vs.\ enclosed mass at
    different times for the model with the W\&H EoS. Times are normalized to
    core bounce.}}\label{promptW}
\end{figure}

% Fig: 
\begin{figure}[!t]{
\resizebox{8.7cm}{!}{
\includegraphics{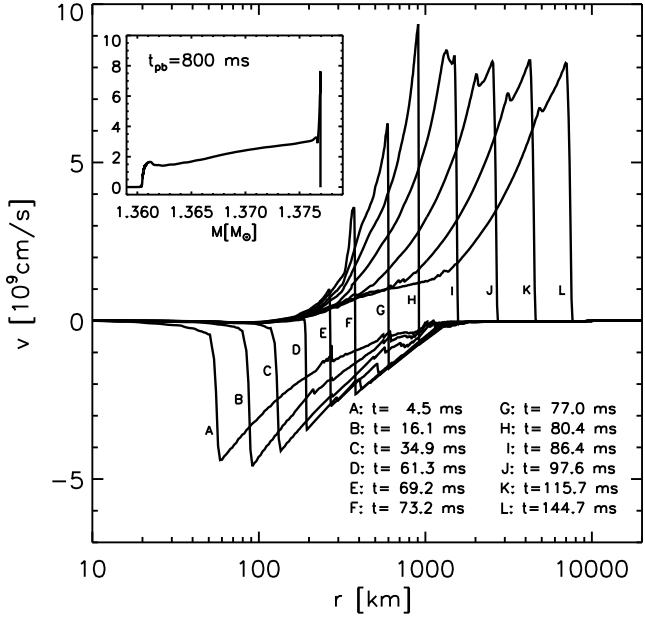}}}
\caption{\footnotesize{Velocity profiles as functions of radius for different
    post-bounce times for the simulation with the W\&H EoS. The insert shows
    the velocity profile vs.~enclosed mass at the end of our simulation.}}\label{vexpbW}
\end{figure}

% Fig: 
\begin{figure}[!t]{
\resizebox{8.7cm}{!}{
\includegraphics{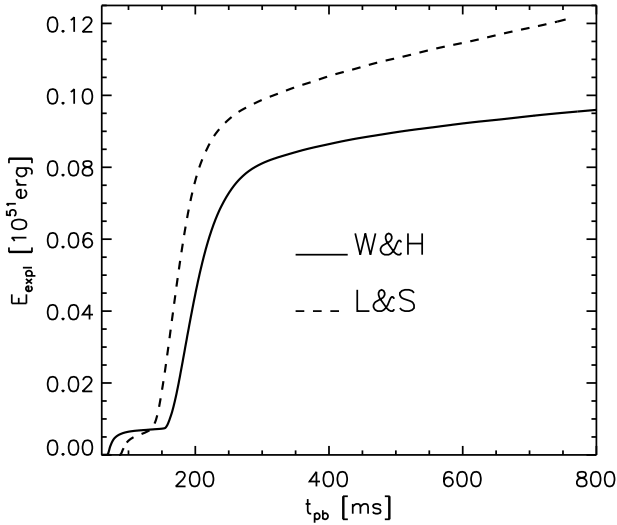}}}
\caption{\footnotesize{Explosion energies as functions of time for the
    simulations with the W\&H and L\&S EoSs. The energy is defined as the
    volume integral of the total gas energy (internal plus kinetic plus
    gravitational) in regions where the latter is positive.}}\label{ener}
\end{figure}

The code is augmented with improved microphysics as described in Buras et
al.~(2005). It includes also the improved treatment of electron captures on a
large variety of nuclei in nuclear statistical equilibrium (NSE), based on shell
 model Monte Carlo calculations, as described by Langanke et al.~(2003). In
 addition, electron captures on certain important nuclei in the non-NSE regime,
 in particular $^{20}$Ne
 and $^{24}$Mg, are implemented according to Takahara et al.~(1989). A
 simplified treatment of  nuclear burning accounts for the main reactions of seven
symmetric nuclei (He, C, O, Ne, Mg, Si, Ni). Details about the implemented
microphysics will be described in a forthcoming paper (Kitaura et al., in
preparation). The nuclear burning reactions considered by Hillebrandt et
al.~(1984) (${\rm^{12}C}{+\rm^{12}C}$, ${\rm^{16}O}{+\rm^{16}O}$,
${\rm^{12}C}{+\rm^{16}O}$) are all included, taking into account different reaction channels.  

% Fig: 
\begin{figure}[!t]{
\resizebox{8.7cm}{!}{
\includegraphics{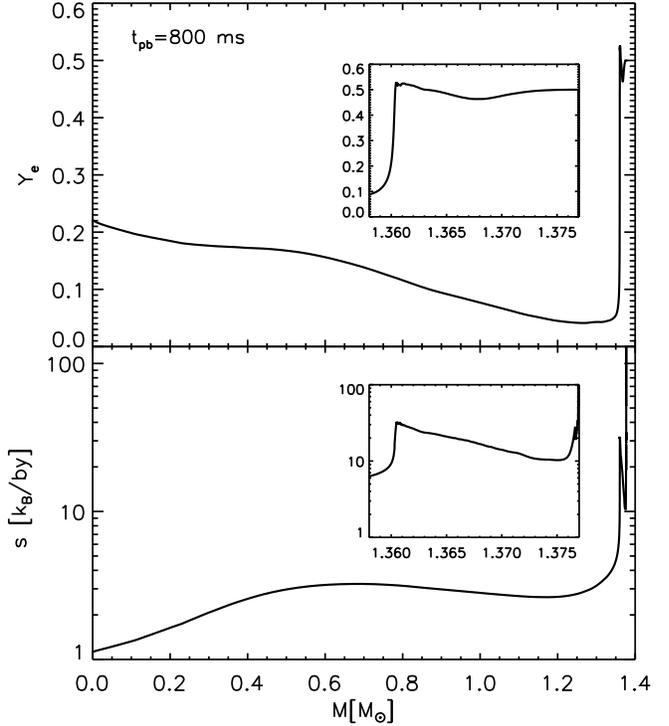}}}
\caption{\footnotesize{Profiles of the electron fraction $Y_{\rm{e}}$ and
    entropy $s$ as functions of enclosed mass at the end of our simulation with
    the W\&H EoS.}}\label{YEWH}
\end{figure}

To describe matter in NSE, we use two different nuclear EoSs in separate
simulations, the Wolff \& Hillebrandt (W\&H) EoS (Hillebrandt et al.\ 1984), 
which is based on Hartree Fock
calculations, and the Lattimer \& Swesty (L\&S) EoS (1991), which is a
finite-temperature compressible liquid-drop model and has a compressibility
modulus of 180 MeV.

This permits us to compare our models with those of Hillebrandt et al.~(1984),
 in which the Wolff and Hillebrandt EoS was used and which produced prompt explosions. 
 The low-temperature and low-density EoS outside of the NSE regime is described
 by an ideal gas of nuclei and nucleons, electrons, positrons, and
photons (Janka 1999). The switch between NSE and non-NSE
 description was made in a density- and temperature-dependent manner.
%{\it The initial data.}
The initial model is the same as the one used in previous SN
calculations of O-Ne-Mg cores by Hillebrandt et al.~(1984) and Mayle \& Wilson (1988). It is a 2.2 $M_\odot$ He core that corresponds
to a progenitor with a main sequence mass of $\sim$8.8$M_\odot$ (Nomoto 1984,
1987). Prior to collapse it has an O-Ne-Mg core with $\sim$1.3$ M_\odot$,
surrounded by a C-O shell of about $0.08 M_\odot$. We take, however, the initial data at a time when the  central density is $\sim$4$\times10^{10}$
g/cm$^3$ and only $\sim$0.1 solar masses at the center of the core
have reached nuclear statistical equilibrium (NSE). This is earlier than the
starting configuration taken by other groups, where the core had already a central density of $\sim$3$\times10^{11}$ g/cm$^3$ and where around 0.3 solar
masses were in NSE. Our earlier initial model allows us to trace the evolution
of the core towards collapse. 

 We added a helium atmosphere of about 10$^{-4} M_\odot$  around the O-Ne-Mg
 and C-O core, so that we could move the outer boundary of
our Eulerian grid from the core radius of about 1100 km to 100 000 km. For the
He-shell profile we adopted a power-law like behaviour of the temperature
($T\propto{r^{-1}}$) from a 10.2 $M_\odot$ progenitor of A.~Heger (private
 communication), and constructed the density profile by assuming hydrostatic
 equilibrium, a mass fraction of 100\% He, and using the
EoS for the low-density regime.
We employ in our simulations a very fine mesh in order to resolve the steep
density gradient at the outer boundary of the C-O layer, with 1600 nonequidistant zones for the hydrodynamics part. The neutrino transport is done with 235 nonequidistant radial zones. 

\section{Results}
Our results show a very smooth transition from
the preceding progenitor evolution to core-collapse. Since
O-Ne-Mg cores are gravitationally less bound than more massive stellar progenitors and can release more energy due to nuclear burning, a temperature-
and density-dependent treatment of all relevant nuclear burning reactions had
to be included, combined with a detailed description of the important electron
capture rates. Only that ensured that the progenitor evolution continued
towards gravitational collapse without numerical transients. We could therefore confirm that the
neutrinos produced by electron captures carry away efficiently the energy that
is released by nuclear burning (Miyaji et al.~1980, Miyaji \& Nomoto 1987, Hashimoto \& Nomoto 1993). A
cruder burning treatment, or omission of
the improved electron capture rates on nuclei, can have the consequence that the
core expands instead of collapsing to a NS, as we verified in test calculations.
  Fig.~\ref{masW} shows how the mass
shells in the inner region during the first milliseconds start
contracting towards the center. The collapse of the core proceeds to higher central
densities and when the density of nuclear matter is reached, the EoS
``stiffens'', and the inner homologous core bounces. The supersonically
falling outer layers collide
with this central core and a hydrodynamic shock is formed as can be seen in
Fig.~1, where this discontinuity in the fluid flow becomes visible by sharp kinks of the mass
shell trajectories. 
 This happens after 59 ms of collapse for the calculation with the W\&H
EoS and
after 78 ms in the calculation with the softer L\&S EoS. The shock
formation radius is at 1.40$\times10^{6}$ cm, corresponding to an enclosed
mass of $0.475 M_\odot$ in the former case, and at
1.15$\times10^{6}$ cm with an enclosed mass of $0.425 M_\odot$ in the
second model. 
 This so-called prompt bounce shock produces initially positive velocities in the
 postshock matter. However, the energy of the shock is insufficient to
 cause a prompt explosion, and the photodisintegration of nuclei consumes such
 amounts of energy that the shock is quickly damped and that within only
 $\sim$1.2 ms after shock formation the velocities are negative everywhere (see
 Fig.~\ref{promptW} for the case of the W\&H EoS). Therefore, the prompt shock
 mechanism fails, independent of the employed nuclear EoS.  

The subsequent expansion of the shock is supported by a combination of different
effects. Initially very high mass accretion rates cause the material to pile up
between neutrinosphere and shock as it is also observed in the early
post-bounce accretion phase of more massive progenitor stars (see, for example,
Buras et al.~2005). Second, the rapid decrease of the mass
accretion rate contributes to ensure ongoing expansion, because even for
quasi-stationary conditions the accretion shock adjusts to a larger radius for smaller mass accretion rates. Finally, as
soon as the shock reaches the outer edge of the C-O shell, a
very steep density decline leads to an outward acceleration of the shock. The last two
aspects are linked to the specific structure of O-Ne-Mg cores and discriminate
SN progenitors with such cores from more massive stars.
However, despite the shock expansion the material behind the shock has
initially still negative velocities and is accreted onto the forming
NS (Figs.~\ref{masW},\ref{vexpbW}). Note that when the matter
right behind the shock starts to expand with the shock, the gas
accreted by the shock is gravitationally bound and remains so in passing
through the shock. However, $p$d$V$ work excerted from below and to a minor
extent energy input by
neutrino heating can convert the accretion
into an explosion, accelerating a tiny amount of matter ($\ll
10^{-3}\,M_\odot$) to move outward with the shock. 

While the shock reaches larger radii, the temperature and density behind the
shock decrease. High-energy electron neutrinos and antineutrinos, which stream
off from their neutrinospheres (represented by the thick solid, dashed, and dash-dotted lines in Fig.~\ref{masW})
begin to deposit energy behind the shock mainly by absorption on nucleons. This
leads to the formation of a ``gain radius'' (thin dashed line in Fig.~\ref{masW}) which separates a layer of neutrino cooling around the neutrinosphere from the
energy ``gain layer'' behind the shock (Bethe \& Wilson 1985).

About 80 ms after bounce for the L\&S EoS and about 60 ms for the W\&H EoS, the neutrino heating timescale, defined by the total energy in the gain layer divided by the neutrino heating rate in that region, gets
smaller than the advection timescale. The latter is given as the time the
accreted matter needs for being advected from the shock to the gain radius. Since the increasing shock
radius leads to smaller and smaller postshock velocities, the duration of the deposition of
energy via neutrino absorption in the shocked matter increases. The continuous input of
energy raises the total energy of the gain layer to a
value near zero within roughly 100 ms, unbinding the matter in the gravitational field of the
forming NS. The
fluid velocity in the layer close to the gain radius therefore starts
to become positive (see Figs.~\ref{masW},\ref{vexpbW}) and the explosion energy
shows a rapid increase (Fig.~\ref{ener}).

 The cooling region then becomes more and more narrow as the gain radius retreats
 towards the neutrinosphere, so that neutrinos diffusing out of the contracting
 protoneutron star begin to heat the layers right above the
 neutrinosphere. Gas is thus ablated from the NS surface, and the so-called neutrino-driven
 wind phase sets in at $t\ga200\,$ms after bounce. Several of the
mass shells depicted in Fig.~\ref{masW} clearly show this process. 

At this time the energy in the expanding postshock matter has increased to about
 $0.1\times10^{51}$ erg, rising further due to the power input by the
 neutrino-driven wind. We extrapolate that the final energy of the explosion
 will be slightly larger than $0.1\times10^{51}$ erg for the calculation with
 the W\&H EoS and  might be about 50\% higher in case of the L\&S EoS. This is
 roughly a factor of 10 lower
 than the canonical SN value, in contrast to the findings in previous explosion
 models of O-Ne-Mg cores (Hillebrandt et al.~1984, Mayle \& Wilson 1988). 

 At the end of our simulation the mass cut, and therefore the baryon mass of the
 protoneutron star, is around 1.360 $M_\odot$ for the W\&S EoS
 (Fig.~\ref{masW}) and about $1.363$ $M_\odot$ for the L\&S EoS. The NS mass
 will only slightly decrease further because of the ongoing mass loss in the
 neutrino-driven wind. The mass of the ejecta lies
 therefore between 0.014 and 0.017 $M_\odot$. The ejected gas has an
 electron fraction, $Y_{\rm{e}}$, between 0.46 and 0.53 and entropy values between 10 and 40
$k_{\rm{B}}$ per nucleon for both EoSs (see
 Fig.~\ref{YEWH} for the case with the W\&H EoS). Since only about one third of the ejected matter
 has a $Y_{\rm{e}}$ value very close to 0.5, the mass of ejected $^{56}$Ni is
 certainly smaller than $\sim$0.015$M_\odot$. 
The $Y_{\rm{e}}$ values in our models are higher than those in previous
 simulations of SNe from O-Ne-Mg cores (Mayle \& Wilson 1988). This points to important differences in the neutrino
 treatment. Spectral Boltzmann transport calculations have recently found early
 ejecta with $Y_{\rm{e}}$ around 0.5 and higher also in (artificial) explosions of more massive
 progenitors (Buras et al.~2005, Fr\"ohlich et al.~2005). The reason for
 this difference compared to the older models is a refined description of
 neutrino spectra formation and in particular of charged-current
 neutrino-nucleon interactions, including the weak magnetism
 corrections that were pointed out to be relevant by Horowitz (2002). 

\section{Conclusions}
Our 1D simulations of SN explosions from the collapse of
O-Ne-Mg cores suggest that such SNe are powered by neutrino heating and by the
neutrino-driven wind of the newly formed NS, similar to what Woosley \& Baron (1992) found in case of the accretion-induced collapse of white dwarfs to
NSs. Such events have a
low explosion energy ($\sim\,$0.1$\times 10^{51}\,$erg)
and produce little $^{56}$Ni ($\la 10^{-2}\,M_\odot$).
Most of the ejecta expand initially with velocities of
2--4$\times 10^4\,$km$\,$s$^{-1}$, a small fraction has
nearly $10^{5}\,$km$\,$s$^{-1}$. This is significantly faster
than in SNe of more massive progenitors. Of course, sweeping up the matter of
the stellar mantle and envelope, the shock will decelerate, and the ejecta
velocities after shock breakout from the stellar surface will be
correspondingly lower. During the first
second of the explosion, the ejected matter has
$0.46\la Y_{\mathrm{e}}\la 0.53$ and modest entropies
($10\la s/(k_{\mathrm{B}}/{\rm{by}}) \la 40$). Such conditions exclude
that r-process elements are formed in this matter during this early phase of
the explosion of O-Ne-Mg cores: $Y_{\mathrm{e}}$ is too large for
the ``classical'' low-entropy r-process and $s$ is too low for
high-entropy r-processing. The ejecta in our models, however,
do not show the vast overproduction of some very neutron-rich,
rare isotopes like $^{87}$Kr, which was made in low-$Y_{\mathrm{e}}$
material ($Y_{\mathrm{e}}\la 0.44$) in previous simulations
(Mayle \& Wilson 1988), and which was interpreted as a severe constraint
to the rate of such events.
Our models yield considerably less energetic explosions than
previous simulations and show other significant differences
in the dynamics and explosion characteristics. These are
probably mainly linked to the improved treatment of neutrino
transport and neutrino-matter interactions. 

The small explosion energy obtained in our simulations is more   
consistent than previous explosion models of O-Ne-Mg cores with
the low present expansion velocities ($\sim\,$1500$\,$km$\,$s$^{-1}$)
of the filaments of the Crab remnant of SN~1054 (Davidson \& Fesen 1985),
corresponding to a low kinetic energy of
0.6--1.5$\times 10^{50}\,$erg for an ejecta mass of
$4.6\pm 1.8\,M_\odot$ in ionized and neutral gas (Fesen et al.\ 1997).
The energy could be even somewhat larger if there were several solar
masses of material in an undetected, extended halo.

While our simulations are spherically symmetric, we do not expect
any qualitative changes in the multi-dimensional case, and
probably only a modest increase of the explosion energy.
Since very fast outflow develops on a relatively short timescale
after core bounce, nonradial hydrodynamic instabilities are 
unlikely to have time to merge and grow to very large structures or 
global asymmetry before the anisotropic pattern freezes out in
the accelerating expansion (Scheck et al.\ 2006, in preparation).
Therefore the recoil velocity of the NS due
to anisotropic mass ejection should remain fairly small (see
Scheck et al.\ 2004), in agreement with speculations by 
Podsiadlowski et al.\ (2004). Corresponding 2D simulations are 
in progress. 

Our models have also an important bearing on the nuclear EoS 
constraints deduced by Podsiadlowski et al.\ (2005) from the 
low-mass Pulsar~B of the double pulsar J0737-3039, which has
a gravitational mass of $M_{\mathrm{G}} = 1.249\pm 0.001\,M_\odot$. 
Provided the 
progenitor model we use is valid, the mass loss of the collapsing
O-Ne-Mg core during the explosion leaves the neutron star with
a baryonic mass of $M_0 = 1.36\pm 0.002\,M_\odot$. The 
error range accounts approximately for
variations associated with the employed EoS and the wind ablation 
after our simulations are terminated. Our value implies a systematic
left shift and reduction of the ``acceptance rectangle'' in Fig.~3 
of Podsiadlowski et al.\ (2005). Combined with the recent
measurement of a pulsar of 2.1$\pm 0.2\,M_\odot$ in 
PSR$\,$J0751+1807 (Nice et al.\ 2005), which is the largest well
determined NS mass so far, this lends viability only to a limited
number of NS EoSs which allow for a sufficiently large maximum mass
and whose $M_{\mathrm{G}}/M_0$-curves pass through the 
acceptance rectangle.

Based on our findings one might speculate that (B-) stars around 
$9\,M_\odot$ are also the progenitors of some of the subluminous Type II-P 
supernovae mentioned in the introduction. 
In fact, their peculiarities would be
explained in a very natural way. The low peak luminosity and extended
plateau phase could result from the combination of a low hydrogen
envelope mass ($\simeq\,$6$\,M_\odot$) with low expansion velocities
($\la\,$3000 km/s). The small mass of radioactive $^{56}$Ni would
explain the low tail-luminosity of these objects.
An alternative interpretation of subluminous Type II-P supernovae is the
explosion of rather massive stars with extended envelopes, but
otherwise more ``normal'' explosion energies. This connection is 
supported by the long duration of the plateau phase of many of 
these events (cf.\ Pastorello et al.\ 2004, and refs.\ therein).
Provided the observations cover the full duration of the plateau, 
SN~1997D and 2003gd may still be viable cases for explosions of stars
with main sequence masses around 9 $M_\odot$ (Hendry et al.\ 2005).
There is a clear difference between such stars and more massive supernova
progenitors. The former eject very little amounts (some $10^{-3}\,M_\odot\,$?)
of oxygen only, like SN~1997D (Chugai \& Utrobin 2000),
whereas the latter produce up to a solar mass or more. Therefore one
might be able to distinguish between the two scenarios on the basis of
observations by measuring the oxygen lines in the late nebular spectra.

{\it Acknowledgements.}
We are grateful to K.~Nomoto for providing us with the initial data and for
encouraging discussions. We thank R.~Buras and M.~Rampp for their
input to this work, A.~Marek especially for the W\&H nuclear EoS, and 
P.~Mazzali,
A.~Pastorello, D.~Sauer and V. Utrobin for very helpful discussions about 
SN observations.
This work was supported by the Sonderforschungsbereich 375 ``Astroparticle
Physics'' of the Deutsche Forschungsgemeinschaft and by the 
International Max Planck Research School (IMPRS).
Supercomputer time at the Rechenzentrum Garching is acknowledged.

\nocite{maywil88}
\nocite{wanom03}
\nocite{whecowhi98}
\nocite{sumter01}
\nocite{hilnom84}
%\nocite{hilwol85}
\nocite{burla85}
\nocite{barcoka87}
\nocite{ramjan02}
\nocite{mnys80}
\nocite{miynom87}
\nocite{nom82}
\nocite{nom83}
\nocite{nom84}
\nocite{nom87}
\nocite{hainom93}
\nocite{frybeheco99}
\nocite{wooba92}
\nocite{betwil85}
\nocite{arn75}
\nocite{gott70}
\nocite{dav82}
\nocite{lan03}
\nocite{tak89}
\nocite{past05}
\nocite{hil78}
\nocite{hil82}
\nocite{woo80}
\nocite{latswe91}
\nocite{ham03}
\nocite{past04}
\nocite{chu00}
\nocite{hor02}
\nocite{hart85}
\nocite{mar05}
\nocite{bur05}
\nocite{jan99}
\nocite{fro04}
\nocite{hen05}
\nocite{zam03}
\nocite{fes97}
\nocite{dav85}

{\small

\bibliographystyle{aa}
\bibliography{lit}

\begin{thebibliography}{38}
\expandafter\ifx\csname natexlab\endcsname\relax\def\natexlab#1{#1}\fi

\bibitem[{{Arnett}(1975)}]{arn75}
{Arnett}, W.D. 1975, \apj, 195, 727

\bibitem[{{Baron} {et~al.}(1987){Baron}, {Cooperstein}, \&
  {Kahana}}]{barcoka87}
{Baron}, E., {Cooperstein}, J., \& {Kahana}, S. 1987, \apj, 320, 300

\bibitem[{{Bethe} \& {Wilson}(1985)}]{betwil85}
{Bethe}, H.A. \& {Wilson}, J.R. 1985, \apj, 295, 14

\bibitem[{{Buras} {et~al.}(2005){Buras}, {Rampp}, {Janka}, \&
  {Kifonidis}}]{bur05}
{Buras}, R., {Rampp}, M., {Janka}, H.-T., \& {Kifonidis}, K. 2005, \aap,
  in the press, astro-ph/0507135

\bibitem[{{Burrows} \& {Lattimer}(1985)}]{burla85}
{Burrows}, A. \& {Lattimer}, J. 1985, \apj, 299, L19

\bibitem[{{Chugai} \& {Utrobin}(2000)}]{chu00}
{Chugai}, N.N. \& {Utrobin}, V.P. 2000, \aap, 354, 557

\bibitem[{{Davidson} \& {Fesen}(1985)}]{dav85}
{Davidson}, K. \& {Fesen}, R.A. 1985, ARAA, 23, 119

\bibitem[{{Davidson, K. et al.}(1982)}]{dav82}
{Davidson, K. et al.} 1982, \apj, 253, 696

\bibitem[{{Fesen} {et~al.}(1997){Fesen}, {Shull}, \&
  {Hurford}}]{fes97}
{Fesen}, R.A., {Shull}, J.M., \& {Hull}, A.P. 1997, \aj, 113, 354

\bibitem[{{Fr\"ohlich, C. et al.}(2004)}]{fro04}
{Fr\"ohlich, C. et al.} 2004, \apj, in the press, astro-ph/0410208

\bibitem[{{Fryer} {et~al.}(1999){Fryer}, {Benz}, {Herant}, \&
  {Colgate}}]{frybeheco99}
{Fryer}, C.L., {Benz}, W., {Herant}, M., \& {Colgate}, S. 1999, \apj, 516, 892

\bibitem[{{Gott} {et~al.}(1970){Gott}, {Gunn}, \& {Ostriker}}]{gott70}
{Gott}, J.R., {Gunn}, J.E., \& {Ostriker}, J.~P. 1970, \apj, 160, L91

\bibitem[{{Hamuy}(2003)}]{ham03}
{Hamuy}, M. 2001, \apj, 582, 905

\bibitem[{{Hartmann} {et~al.}(1985){Hartmann}, {Woosley}, \& {El Eid}}]{hart85}
{Hartmann}, D., {Woosley}, S.E., \& {El Eid}, M.F. 1985, \apj, 297, 837

\bibitem[{{Hashimoto} {et~al.}(1993){Hashimoto}, {Iwamoto}, \&
  {Nomoto}}]{hainom93}
{Hashimoto}, M., {Iwamoto}, K., \& {Nomoto}, K. 1993, \apj, 414, L105

\bibitem[{{Hendry et al.}(2005)}]{hen05}
{Hendry, M.A. et al.} 2005, \mnras, 359, 906

\bibitem[{{Hillebrandt}(1978)}]{hil78}
{Hillebrandt}, W. 1978, Space Science Reviews, 21, 639

\bibitem[{{Hillebrandt}(1982)}]{hil82}
{Hillebrandt}, W. 1982, \aap, 110, L3

\bibitem[{{Hillebrandt} {et~al.}(1984){Hillebrandt}, {Nomoto}, \&
  {Wolff}}]{hilnom84}
{Hillebrandt}, W., {Nomoto}, K., \& {Wolff}, R.G. 1984, \aap, 133, 175

% \bibitem[{{Hillebrandt} \& {Wolff}(1985)}]{hilwol85}
% {Hillebrandt}, W. \& {Wolff}, R. 1985, in {Nucleosynthesis --- Challenges and
%   New Developments}, ed. W.D. {Arnett} \& J.W. {Truran} (Chicago: University
%  of Chicago Press), p.~131

\bibitem[{{Horowitz}(2002)}]{hor02}
{Horowitz}, C.J. 2002, \prd, 65, 043001

\bibitem[{{Janka}(1999)}]{jan99}
{Janka}, H.-T. 1999, unpublished

\bibitem[{{Langanke} {et~al.}(2003){Langanke}, {Martinez-Pinedo}, {Sampaio},
  {Dean}, {Hix}, {Messer}, {Mezzacappa}, {Liebend\"orfer}, {Janka}, \&
  {Rampp}}]{lan03}
{Langanke}, K., {Martinez-Pinedo}, G., {Sampaio}, J.M., {et~al.} 2003, \prl,
  90, 241102

\bibitem[{{Lattimer} \& {Swesty}(1991)}]{latswe91}
{Lattimer}, J. \& {Swesty}, F. 1991, Nucl.~Phys.~A, 535, 331

\bibitem[{{Marek} {et~al.}(2005){Marek}, {Dimmelmeier}, {Janka}, {M\"uller}, \&
  {Buras}}]{mar05}
{Marek}, A., {Dimmelmeier}, H., {Janka}, H.-T., {M\"uller}, E., \& {Buras}, R.
  2005, \aap, in the press, astro-ph/0502161

\bibitem[{{Mayle} \& {Wilson}(1988)}]{maywil88}
{Mayle}, R. \& {Wilson}, J.R. 1988, \apj, 334, 909

\bibitem[{{Miyaji} \& {Nomoto}(1987)}]{miynom87}
{Miyaji}, S. \& {Nomoto}, K. 1987, \apj, 318, 307

\bibitem[{{Miyaji} {et~al.}(1980){Miyaji}, {Nomoto}, {Yokoi}, \&
  {Sugimoto}}]{mnys80}
{Miyaji}, S., {Nomoto}, K., {Yokoi}, K., \& {Sugimoto}, D. 1980, \asj, 32, 303

\bibitem[{{Nice} {et~al.}(2005){Nice}, {Splaver}, {Stairs}, {L\"ohmer}, 
  {Jessner}, {Kramer}, \& {Cordes}}]{nice05}
{Nice}, D.J., {Splaver}, E.M., {Stairs}, I.H., {et~al.} 2005, \apj, 
  submitted, astro-ph/0508050

\bibitem[{{Nomoto}(1983)}]{nom83}
{Nomoto}, K. 1983, \iau, 101, 139N

\bibitem[{{Nomoto}(1984)}]{nom84}
{Nomoto}, K. 1984, \apj, 277, 791

\bibitem[{{Nomoto}(1987)}]{nom87}
{Nomoto}, K. 1987, \apj, 322, 206

\bibitem[{{Nomoto} {et~al.}(1982){Nomoto}, {Sparks}, {Fesen}, {Gull}, {Miyaji},
  \& {Sugimoto}}]{nom82}
{Nomoto}, K., {Sparks}, W.M., {Fesen}, R.A., {et~al.} 1982, \nat, 299, 803

\bibitem[{{Pastorello, A. et al.}(2004)}]{past04}
{Pastorello, A. et al.} 2004, MNRAS, 347, 74

\bibitem[{{Pastorello, A. et al}(2005)}]{past05}
{Pastorello, A. et al}. 2005, MNRAS, submitted

\bibitem[{{Podsiadlowski} {et~al.}(2004){Podsiadlowski}, {Langer}, 
{Poelarends}, {Rappaport}, {Heger}, \& {Pfahl}}]{pod04} 
{Podsiadlowski}, Ph., {Langer}, N., {Poelarends}, A.J.T., {et~al.} 2004, 
 \apj, 612, 1044

\bibitem[{{Podsiadlowski} {et~al.}(2005){Podsiadlowski}, {Dewi}, 
{Lesaffre}, {Miller}, {Newton}, \& {Stone}}]{pod05}
{Podsiadlowski}, Ph., {Dewi}, J.D.M., {Poelarends}, P., {et~al.} 2005, 
 MNRAS, 361, 1243

\bibitem[{{Rampp} \& {Janka}(2002)}]{ramjan02}
{Rampp}, M. \& {Janka}, H.-T. 2002, \aap, 396, 361

\bibitem[{{Scheck} {et~al.}(2004){Scheck}, {Plewa}, {Janka}, {Kifonidis}, 
  \& {M\"uller}}]{mul04}
{Scheck}, L., {Plewa}, T., {Janka}, H.-T., {et~al.} 2004, PRL, 92, 011103

\bibitem[{{Sumiyoshi} {et~al.}(2001){Sumiyoshi}, {Terasawa}, {Mathews},
  {Kajino}, {Yamada}, \& {Suzuki}}]{sumter01}
{Sumiyoshi}, K., {Terasawa}, M., {Mathews}, G.J., {et~al.} 2001, \apj, 562,
  880

\bibitem[{{Takahara} {et~al.}(1989){Takahara}, {Hino}, {Oda}, {Muto},
  {Wolters}, \& {Glaudemans}}]{tak89}
{Takahara}, M., {Hino}, T., {Oda}, T., {et~al.} 1989, Nucl. Phys.~A, 504, 167

\bibitem[{{Wanajo} {et~al.}(2003){Wanajo}, {Tamamura}, {Naoki}, {Nomoto},
  {Yuhri}, {Beers}, \& {Nozawa}}]{wanom03}
{Wanajo}, S., {Tamamura}, M., {Naoki}, I., {et~al.} 2003, \apj, 593, 968

\bibitem[{{Wheeler} {et~al.}(1997){Wheeler}, {Cowan}, \&
  {Hillebrandt}}]{whecowhi98}
{Wheeler}, J.C., {Cowan}, J.J., \& {Hillebrandt}, W. 1997, \apj, 493, L101

\bibitem[{{Woosley} \& {Baron}(1992)}]{wooba92}
{Woosley}, S.E. \& {Baron}, E. 1992, \apj, 391, 228

\bibitem[{{Woosley} {et~al.}(1980){Woosley}, {Weaver}, \& {Taam}}]{woo80}
{Woosley}, S.E., {Weaver}, T.A., \& {Taam}, R. 1980, in {Type I Supernovae},
  ed. J.C. {Wheeler} (Austin: University of Texas), p.~96

\bibitem[{{Zampieri et al.}(2003)}]{zam03}
{Zampieri, L. et al.} 2003, \mnras, 338, 711

\end{thebibliography}

}
  
\end{document}